%% file: main.tex
\documentclass[10pt,conference]{IEEEtran}
\IEEEoverridecommandlockouts
\usepackage{algorithmic}
\usepackage{graphicx}
\usepackage{stfloats}%for correct floating of two-columns figures
\usepackage{textcomp}
\usepackage{xcolor}
\usepackage{xspace}
\usepackage[nointegrals]{wasysym}
\usepackage{pifont}
\usepackage{xcolor}
\usepackage{multirow}
\usepackage{flushend}
\usepackage{caption}
\usepackage{booktabs}
\usepackage{algorithm}
\usepackage{algorithmic}
\usepackage{pdfpages}
\usepackage{amssymb}

\usepackage[hyphens]{url} % 支持网址自动换行
\usepackage{hyperref} % 生成可点击的超链接
\hypersetup{breaklinks=true} % 允许链接断行

\usepackage{tabularx}
\usepackage{multicol} %Jianjun Zhao 20241009
\usepackage[inkscapelatex=false]{svg}
\hypersetup{
colorlinks = true, % false: boxed links; true: colored links
linkcolor=black, % color of internal links
citecolor=black, % color of links to bibliography
urlcolor=black % color of external links
}
\usepackage{url}
\usepackage{enumitem}
\usepackage{braket}
\usepackage{amsmath}
\usepackage{caption}
\usepackage[breakable]{tcolorbox}
\usepackage{colortbl}
\usepackage{pifont}
\usepackage{float}                  % 图片浮动位置
\usepackage{subfig}                 % 子图包，不要与{subfigure}混用，{subfig}较新
\usepackage{overpic}                % 叭啦叭啦，与图片排版相关

\usepackage{tikz}
\usepackage{color}
\usepackage{listings}
\input{macros}
\begin{document}
%
% --- Author Metadata here ---
%\conferenceinfo{NIME'17,}{May 15-19, 2017, Aalborg University Copenhagen, Denmark.}
\title{Unveiling Code Clones in Quantum Programming: An Empirical Study with Qiskit}

\author{\IEEEauthorblockN{Kenta Manoku}
\IEEEauthorblockA{%\textit{dept. name of organization (of Aff.)} \\
%\textit{Institute of High Energy Physics}\\
\textit{Kyushu University, Fukuoka, Japan}\\
manoku.kenta.801@s.kyushu-u.ac.jp}
\and
\IEEEauthorblockN{Jianjun Zhao}
\IEEEauthorblockA{%\textit{Faculty of Electrical Engineering and Information Science} \\
%\textit{Department of Information Science and Technology}\\
\textit{Kyushu University, Fukuoka, Japan}\\
zhao@ait.kyushu-u.ac.jp}
}
%\end{comment}

\maketitle

\begin{abstract}
Code clones, referring to identical or similar code fragments, have long posed challenges in classical programming, impacting software quality, maintainability, and scalability. However, their presence and characteristics in quantum programming remain unexplored. This paper presents an empirical study of code clones in quantum programs, specifically focusing on software developed using the Qiskit framework. We examine the existence, distribution, density, and size of code clones in quantum software, revealing a high density of Type-2 and Type-3 clones involving minor modifications. Our findings suggest that these clones are more frequent in quantum software, likely due to the complexity of quantum algorithms and their integration with classical logic. This highlights the need for advanced clone detection and refactoring tools specifically designed for the quantum domain to improve software maintainability and scalability. 
We also discuss the implications of our results for quantum software development and propose future research directions.
\end{abstract}


% \begin{CCSXML}
% <ccs2012>
%    <concept>
%        <concept_id>10011007.10011074</concept_id>
%        <concept_desc>Software and its engineering~Software creation and management</concept_desc>
%        <concept_significance>500</concept_significance>
%        </concept>
%  </ccs2012>
% \end{CCSXML}

% \ccsdesc[500]{Software and its engineering~Software creation and management}

% % this line creates the CCS Concepts section.
% % \printccsdesc

% \keywords{Quantum Computing, Concolic Testing, Test Coverage}

\section{Introduction}\label{sec:intro}
Code clones, defined as identical or highly similar code fragments within a program, have been extensively studied in classical programming~\cite{roy2009comparison,rattan2013software}. These clones pose challenges for software maintenance, mainly when changes made to one fragment are not consistently applied to others, potentially leading to bugs or functionality inconsistencies~\cite{kim2005empirical}. Various clone detection and refactoring techniques have been developed in classical programming to address these issues, thereby improving software quality, maintainability, and scalability~\cite{duala2008clonetracker,rattan2013software}. However, despite the substantial progress in classical clone research, studies focusing on code clones in quantum programming remain unexplored, even as the importance of quantum software continues to grow.

Quantum computing, leveraging principles such as superposition and entanglement, can potentially solve complex problems beyond the reach of classical computers~\cite{barends2014superconducting}. Fields like artificial intelligence~\cite{dunjko2018machine}, computational chemistry~\cite{mcardle2020quantum,cao2019quantum}, and drug design~\cite{zhou2010quantum,raha2007role} could benefit significantly from quantum algorithms~\cite{shor1994algorithms,grover1996fast,harrow2009quantum}. As quantum programming evolves, the increasing size and complexity of quantum programs pose new challenges for code quality management, including detecting and managing code clones.

Unlike classical programming, quantum programming operates within a distinct computational paradigm, involving operations on quantum bits (qubits) that exhibit fundamentally non-classical behavior~\cite{nielsen2010quantum}. This paradigm shift introduces unique challenges in software development~\cite{zhao2020quantum}, such as debugging, optimization, and ensuring the correctness of quantum circuits. Prior work by Jhaveri {\it et al.}~\cite{jhaveri2023cloning} explored a novel approach for code clone detection by expressing it as a subgraph isomorphism problem solved using quantum annealing. While this represents an important first step, it focuses on the application of quantum computing to classical software problems rather than investigating clones specifically within quantum programs.

This study empirically investigates code clones in quantum programming using Qiskit, a widely adopted open-source quantum computing framework~\cite{gadi_aleksandrowicz_2019_2562111}. We aim to detect and characterize these clones, analyzing their existence, distribution density, and size. The key objectives are to (1) identify the presence of code clones in quantum programs, (2) analyze their structural characteristics, and (3) evaluate their impact on the maintainability, scalability, and development of quantum software. Through this investigation, we seek to provide insights for improving quantum program quality as the field moves toward broader industrial deployment.

The rest of this paper is structured as follows: Section~\ref{sec:methodlogy} outlines the methodology for detecting and analyzing code clones in quantum programs. Section~\ref{sec:experimental results} presents the experimental results. Section~\ref{sec:discussion} discusses the implications of these findings for quantum software development. Finally, Section~\ref{sec:conclusion} concludes with future research directions.

\section{Methodology}
\label{sec:methodlogy} 
This section outlines the methodology used to detect and analyze code clones in quantum programming. 
The process includes selecting suitable repositories, detecting code clones, and analyzing the detected clones in detail.

\subsection{Repository Selection Criteria} 
The target programs for this study were GitHub repositories utilizing the Qiskit framework~\cite{gadi_aleksandrowicz_2019_2562111}. Repositories that were part of the Qiskit organization (i.e., those belonging to the Qiskit Organization on GitHub~\cite{qiskit-github}) were excluded to focus on third-party projects. Specifically, we targeted repositories containing the terms \texttt{from qiskit import} or \texttt{import qiskit} in their source code to ensure that the repository actually used the Qiskit library.

\subsubsection{Qiskit Project Collection} 
The repositories were collected using the \texttt{GitHub REST API}~\cite{github-rest-api}. The \texttt{GitHub REST API} enables querying datasets based on specific criteria, but it limits results to the first 1000 matches per query~\cite{github-rest-api-1000-limit}. We used incremental queries with sorting parameters like \texttt{updated\_at} or \texttt{created\_at} to gather a broader set of repositories. 

We searched for repositories containing 'Qiskit' in their names using the \texttt{search repository}~\cite{search-repositories} function, which supports sorting. After cloning these repositories locally, we identified files containing \texttt{from qiskit import} or \texttt{import qiskit} in their source code. A total of 509 Python repositories were collected as of February 2024.

\subsubsection{Selection of Target Programs} 
After cloning the initial set of 509 repositories, we further refined our selection. From these locally cloned repositories, we selected those containing \texttt{from qiskit import} or \texttt{import qiskit} in their source code, resulting in 438 Python repositories. To minimize bias during the analysis, vast repositories that could skew the results were excluded. This led to a final set of 375 Python repositories for the code clone detection phase, as detailed in Section~\ref{sec:experimental results}.

\subsection{Code Clone Detection} 
The clone detection process involves multiple steps, focusing on identifying code clones at the function and class levels within the source code. Each step is outlined below:

\begin{itemize}[leftmargin=1.3em]
\setlength{\itemsep}{1.5pt}
    \item \textit{File Scanning:} The program recursively scans specified directories to identify Python files, storing their paths in an array for subsequent processing. This allows for an exhaustive search of all Python source files within a repository.

    \item \textit{File Reading:} Using the array of file paths, the program reads the contents of each Python file as plain text. 
    %For Jupyter Notebooks, which store code in JSON format, the program extracts code cells' content, ensuring that code fragments embedded within the notebooks are included in the analysis.

    \item \textit{Code Section Extraction:} Code clones are detected at the function and class definition levels. An Abstract Syntax Tree (AST) is employed to parse the source code into function and class units. Specifically, a \texttt{CodeExtractor} class is implemented, inheriting from \texttt{ast.NodeVisitor}~\cite{node-visitor}. The methods \texttt{visit\_FunctionDef} and \texttt{visit\_ClassDef} are overridden to extract code sections during AST traversal. This method allows for precise extraction of function and class definitions, storing them as individual segments for further similarity analysis.

    \item \textit{Code Clone Detection:} The extracted code segments are compared for similarity using the \texttt{SequenceMatcher} class from Python’s \texttt{difflib} module~\cite{sequence-matcher}. This module calculates similarity while ignoring non-semantic elements such as spaces, comments, and empty lines. The similarity calculation~\cite{rattan2013software} is based on gestalt pattern matching, defined as follows:
    \begin{displaymath} 
    \text{Similarity} = \frac{2 \times \text{Number of Matching Characters}}{\text{Length of String 1} + \text{Length of String 2}} 
    \end{displaymath}
    
    A similarity score above a specified threshold classifies two code sections as clones. Type-1 clones are exact matches, while Type-2 and Type-3 clones include minor modifications or changes in identifiers.
\end{itemize}

%%%%%%%%%%%%%%%%%%%%%%%%%%%%%%%%%%%%
\section{Experimental Results}
\label{sec:experimental results}
This section presents the results of code clone detection for the selected quantum programming repositories.

\subsection{Experimental Setting}
Our development environment was set up using a Linux environment on Windows 11 Pro (version 23H2) via the Windows Subsystem for Linux (WSL2). The Linux distribution used was Ubuntu 22.04.2 LTS. Development was conducted using Visual Studio Code (version 1.86.1) with Python (version 3.10.13).

\subsection{Existence of Code Clones} 
Using the method described in Section~\ref{sec:methodlogy}, we investigated the presence of Type-1, Type-2, and Type-3 code clones in quantum programming. Type-1 code clones are exact copies created by copy-and-paste without any modifications. Type-2 clones involve changes in identifiers, while Type-3 clones include more extensive modifications, additions, or deletions. Any detected clone with a similarity score below one is considered a Type-2 or Type-3 code clone.

The analysis revealed that quantum programs contain both Type-1 and Type-2/Type-3 code clones, suggesting that developers frequently copy and adapt code segments. This is possibly driven by the complexity of quantum algorithms and the need to integrate them with classical control logic, making minor modifications as they reuse existing code.

\begin{figure}[t] \centerline{\includegraphics[width=1.0\linewidth]{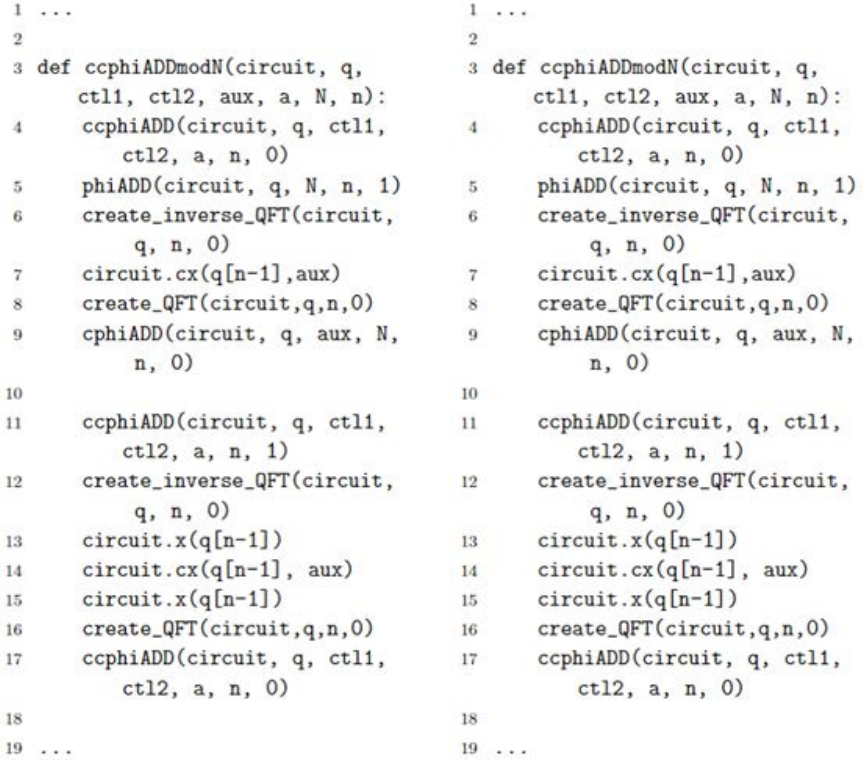}} \caption{An example of a Type-1 code clone is shown. The code on the left is from \small{\texttt{\url{https://github.com/tiagomsleao/ShorAlgQiskit/blob/master/Shor_Normal_QFT.py}}}, while the code on the right is from \small{\texttt{\url{https://github.com/tiagomsleao/ShorAlgQiskit/blob/master/Shor_Sequential_QFT.py}}}.} 
\label{fig:code clone-1}
\end{figure}

\subsection{Distribution and Density of Code Clones} 
\begin{figure*}[t]
%\centering{
	\begin{minipage}[t]{0.45\linewidth}
		\includegraphics[width=3.5in]{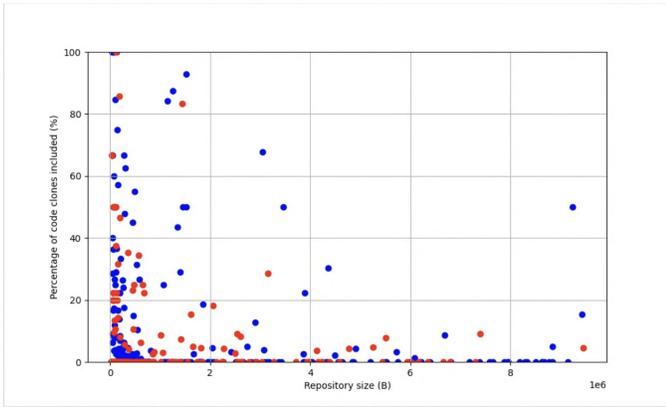}\\
        \centering{(a) Type-1}
	\end{minipage} \hspace*{9.2mm}
	\begin{minipage}[t]{0.45\linewidth}
		\includegraphics[width=3.5in]{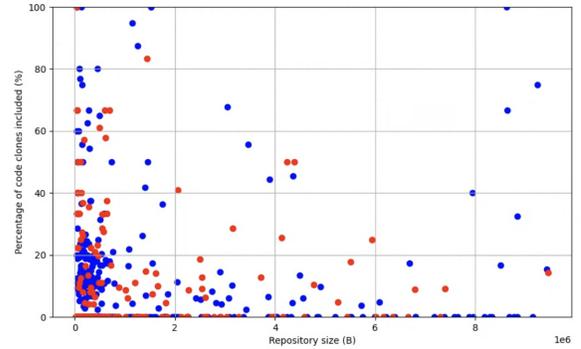}
        \centering{(b) Type-2 and Type-3}
	\end{minipage}
	\caption{Repository size and percentage of files containing code clones}
	\label{fig:hor_2figs_1cap-1}
%}
\end{figure*}

To investigate the distribution and density of code clones, we generated scatter plots that display repository size (in bytes) on the x-axis and the percentage of files containing code clones relative to the total number of files in each repository on the y-axis. As shown in Figures~\ref{fig:hor_2figs_1cap-1}, these scatter plots were generated separately for both Type-1 and Type-2/Type-3 clones. The blue dots represent code clones in classical programs, while the red ones represent quantum programs' code clones. 

\subsubsection{Type-1 Code Clones} The analysis found Type-1 clones distributed across repositories of various sizes. Approximately 13.6\% of the repositories contained Type-1 clones, indicating that exact code duplication remains a common issue even in quantum software. This suggests that developers may be reusing code without modification for consistent quantum circuit operations, which could simplify debugging but pose risks for maintenance. Figure~\ref{fig:code clone-1} shows an example of a Type-1 code clone from the repository %\texttt{ttlion/ShorAlgQiskit}~\cite{ttlion/ShorAlgQiskit}. 
\texttt{\href{https://github.com/tiagomsleao/ShorAlgQiskit/}{tiagomsleao/ShorAlgQiskit}}~\cite{ttlion/ShorAlgQiskit}.

However, the density of Type-1 clones is generally lower in larger repositories, possibly because these repositories may have more mature codebases where direct duplication is minimized through refactoring practices.

\begin{table}[h]
\begin{center}
\caption{Distribution and Density of Code Clones in Quantum Programming (CC: Code Clones)}
Distribution and Density of Type-1 Code Clones
\scalebox{0.75}[0.75]{\begin{tabular}{l|ccccc}
\hline
Size & Total & Repositories & Percentage of & Average Percentage \\
Interval & Repositories & with CC & Repositories with CC & of Files with CC \\
\hline\hline
0 - 53000 & 148 & 3 & 2.03 & 51.11 \\
53000 - 101000 & 66 & 7 & 10.61 & 20.80 \\
101000 - 194000 & 32 & 8 & 25.00 & 45.16 \\
194000 - 370000 & 33 & 4 & 12.12 & 23.88 \\
370000 - 708000 & 32 & 8 & 25.00 & 18.86 \\
708000 - 1353000 & 15 & 4 & 26.67 & 4.35 \\
1353000 - 2586000 & 20 & 9 & 45.00 & 16.67 \\
2586000 - 4944000 & 17 & 4 & 23.53 & 11.24 \\
4944000 - inf & 12 & 4 & 33.33 & 6.56 \\
\hline
All Intervals & 375 & 51 & 13.60 & 9.69 \\
\hline
\end{tabular}}

%%%%%%%%%%%%%%%%%%%%%%%%%%%%%%%%%%%%%%%%%%%%
\vspace*{8mm}
Distribution and Density of Type-2 / Type-3 Code Clones
\scalebox{0.75}[0.75]{\begin{tabular}{l|ccccc}
\hline
Size & Total & Repositories & Percentage of & Average Percentage \\
Interval & Repositories & with CC & Repositories with CC & of Files with CC \\
\hline\hline
0 - 53000 & 148 & 5 & 3.38 & 57.78 \\
53000 - 101000 & 66 & 9 & 13.64 & 29.08 \\
101000 - 194000 & 32 & 9 & 28.12 & 31.43 \\
194000 - 370000 & 33 & 9 & 27.27 & 13.53 \\
370000 - 708000 & 32 & 16 & 50.00 & 31.81 \\
708000 - 1353000 & 15 & 5 & 33.33 & 10.39 \\
1353000 - 2586000 & 20 & 11 & 55.00 & 20.38 \\
2586000 - 4944000 & 17 & 7 & 41.18 & 26.18 \\
4944000 - inf & 12 & 6 & 50.00 & 13.31 \\
\hline
All Intervals & 375 & 77 & 20.53 & 19.17 \\
\hline
\end{tabular}}
\end{center}
\end{table}

\subsubsection{Type-2 and Type-3 Code Clones} 
\begin{figure}[t] \centerline{\includegraphics[width=1.0\linewidth]{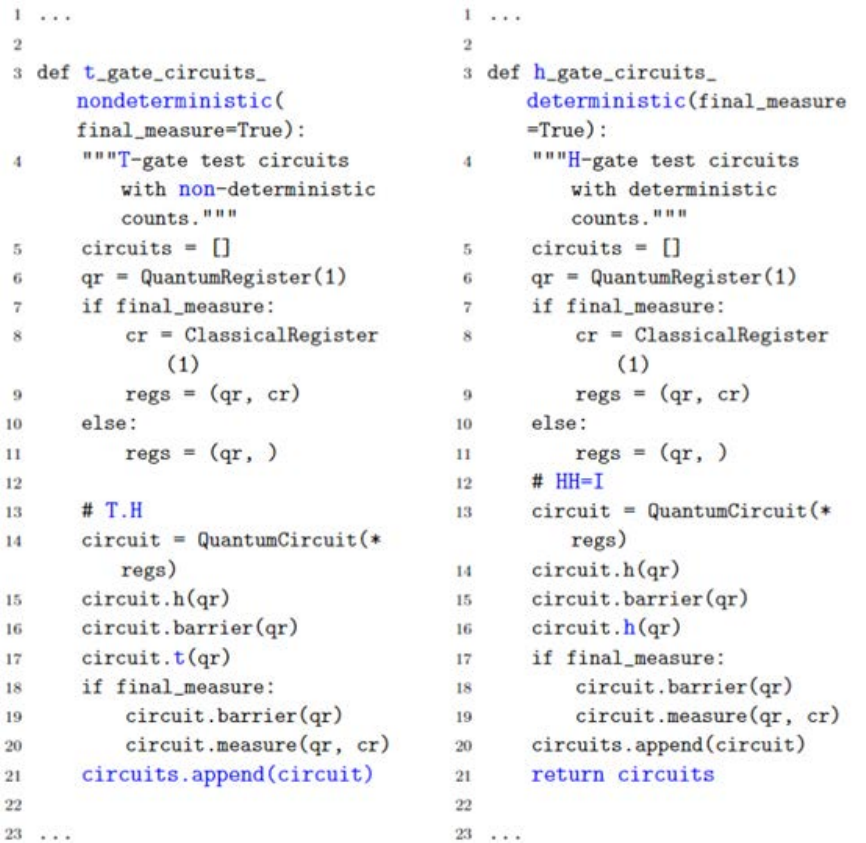}} \caption{An example of a Type-3 code clone is shown. The code on the left is from \small{\texttt{\url{https://github.com/vm6502q/qiskit-qrack-provider/blob/master/test/terra/reference/ref_non_clifford.py}}}, while the code on the right is from \small{\texttt{\url{https://github.com/vm6502q/qiskit-qrack-provider/blob/master/test/terra/reference/ref_1q_clifford.py}}}.} 
\label{fig:code clone}
\end{figure}

Type-2 and Type-3 clones, which involve varying degrees of modification, are more prevalent, with approximately 20.53\% of repositories containing such clones. The higher prevalence suggests that quantum developers often adjust existing code to fit specific needs, such as modifying qubit interactions or optimizing quantum circuit parameters. The increased density of Type-2 and Type-3 clones in smaller repositories may indicate early-stage projects where rapid prototyping leads to frequent minor modifications. In comparison, larger repositories might contain more refactored, stable code with fewer variations in cloned segments.
Figure~\ref{fig:code clone} shows an example of a Type-3 code clone from the repository \texttt{\href{https://github.com/vm6502q/qiskit-qrack-provider/}{vm6502q/qiskit-qrack-provider}}~\cite{vm6502q/qiskit-qrack-provider}. The blue text highlights the differences between the code fragments. 
Although some parts have been edited, most of the code is identical. Through this study, we confirmed that quantum programs also contain both Type-1 and Type-2 / Type-3 code clones.

\subsection{Size of Code Clones} We further analyzed the sizes of the detected code clones. Table~\ref{table:size} provides the number of detected clones and their maximum, minimum, and average sizes for each size interval. The average size of Type-1 code clones in quantum programming was 43 tokens, suggesting that quantum developers may duplicate small functional units or utility functions. In contrast, the average size of Type-2 and Type-3 clones was 26 tokens, indicating that adjustments and small edits to existing code are common practice in quantum programming.

\begin{figure*}
%\centering{
	\begin{minipage}[t]{0.45\linewidth}
		\includegraphics[width=3.5in]{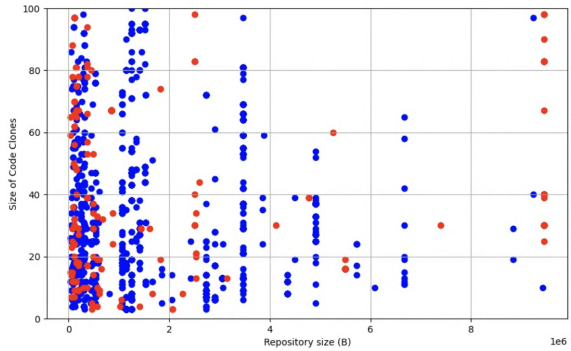}
          \centering{(a) Type-1}
	\end{minipage}\hspace*{9.2mm}
	\begin{minipage}[t]{0.45\linewidth}
		\includegraphics[width=3.5in]{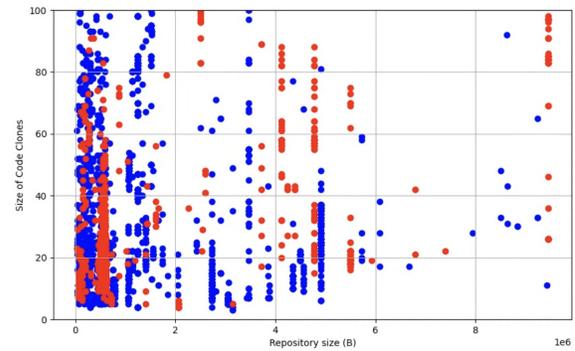}
          \centering{(b) Type-2 and Type-3}
	\end{minipage}
	\caption{Repository size and size of code clones}
	\label{fig:hor_2figs_1cap}
%}
\end{figure*}

Figures~\ref{fig:hor_2figs_1cap} illustrate the relationship between repository size and code clone size. The blue dots represent code clones in classical programs, while the red ones represent quantum programs' code clones. The analysis shows that quantum developers often reuse small code snippets with slight modifications, reflecting a need for precise adjustments in quantum algorithms. For example, changes in quantum circuit parameters or slight variations in qubit operations are frequent. The frequent adjustments might also reflect the iterative nature of developing and optimizing quantum circuits, where developers tweak code to achieve better performance or adapt to different hardware backends.

\begin{table}[h]
\begin{center}
\caption{Sizes of Code Clones in Quantum Programming}
\label{table:size}
Size of Type-1 Code Clones
\scalebox{0.9}[0.9]{\begin{tabular}{l|ccccc}
\hline
Size & Total & Total & Max & Min & Average \\
Interval & Repositories & Code Clones & Size & Size & Size \\
\hline\hline
0 - 53000 & 148 & 4 & 65 & 7 & 36.5 \\
53000 - 101000 & 66 & 13 & 88 & 7 & 31.15 \\
101000 - 194000 & 32 & 126 & 434 & 7 & 58.64 \\
194000 - 370000 & 33 & 221 & 94 & 9 & 13.38 \\
370000 - 708000 & 32 & 31 & 434 & 3 & 57.35 \\
708000 - 1353000 & 15 & 70 & 67 & 4 & 64.14 \\
1353000 - 2586000 & 20 & 35 & 407 & 3 & 115.69 \\
2586000 - 4944000 & 17 & 7 & 126 & 13 & 47.14 \\
4944000 - inf & 12 & 49 & 145 & 16 & 43.57 \\
\hline
All Intervals & 375 & 556 & 434 & 3 & 42.59 \\
\hline
\end{tabular}}

%%%%%%%%%%%%%%%%%%%%%%%%%%%%%%%%%%%%%%%%%%%%%%%%%%%%%%%%
\vspace*{8mm}
Size of Type-2 / Type-3 Code Clones\\
\scalebox{0.9}[0.9]{\begin{tabular}{l|ccccc}
\hline
Size & Total & Total & Max & Min & Average \\
Interval & Repositories & Code Clones & Size & Size & Size \\
\hline\hline
0 - 53000 & 148 & 13 & 191 & 22 & 101.85 \\
53000 - 101000 & 66 & 342 & 190 & 7 & 12.32 \\
101000 - 194000 & 32 & 84 & 127 & 6 & 40.75 \\
194000 - 370000 & 33 & 31 & 160 & 9 & 51.87 \\
370000 - 708000 & 32 & 4763 & 631 & 6 & 22.02 \\
708000 - 1353000 & 15 & 18 & 212 & 5 & 58.33 \\
1353000 - 2586000 & 20 & 81 & 381 & 4 & 77.49 \\
2586000 - 4944000 & 17 & 130 & 2083 & 5 & 92.24 \\
4944000 - inf & 12 & 142 & 294 & 16 & 77.44 \\
\hline
All Intervals & 375 & 5604 & 2083 & 4 & 26.01 \\
\hline
\end{tabular}}
\end{center}
\end{table}

\section{Discussion}\label{sec:discussion}
In this section, we reflect on the findings of the code clone analysis and their implications for quantum programming. 
The results highlight several important aspects of code clones in this context.

\subsection{Implications for Maintenance} Code clones pose well-known challenges to software maintenance, as they can lead to duplicated bugs and inconsistencies when code is modified. In quantum programming, the presence of smaller and more frequent Type-2 and Type-3 clones suggests that developers may need to pay closer attention to maintaining consistency across similar code fragments. Given the intricacies of quantum algorithms, where even slight modifications can lead to significant changes in program behavior, the maintenance of cloned code becomes even more critical.

The density of Type-2 and Type-3 clones in quantum programs indicates potential vulnerabilities, where small, frequently modified code segments may become sources of bugs or inconsistencies. This emphasizes the need for more sophisticated clone detection and refactoring tools tailored specifically to the needs of quantum software development.

\subsection{Challenges in Quantum Software Development} The findings suggest that quantum developers often copy and modify smaller code segments, likely due to the complexity of quantum algorithms and the need to integrate them with classical control logic. This behavior reflects the unique nature of quantum programming, where operations are performed on qubits, often requiring updates to specific code sections. Such a pattern may point to the need for more modular and adaptable code structures in quantum software, as well as tools that can assist in managing these frequent modifications efficiently.

The study also raises questions about the maturity of current quantum programming practices. The prevalence of code clones, especially of Type-2 and Type-3, could indicate that developers are still exploring best practices for code reuse and modularity in this evolving field. Addressing these challenges will be crucial as quantum software scales and matures, moving towards more robust and maintainable development practices.

\section{Conclusion}\label{sec:conclusion}
This paper presents an empirical study on the presence and characteristics of code clones in quantum programming, focusing on software developed using the Qiskit framework. Our findings reveal a notable density of Type-2 and Type-3 clones, characterized by slight modifications or renamed variables. This is likely due to the complexity of quantum algorithms and their integration with classical control logic. Type-1 clones appear at similar rates in both quantum and classical programming.

The frequent occurrence of Type-2 and Type-3 clones in quantum programs poses unique challenges for software maintenance. Even minor modifications can significantly impact the behavior of quantum programs. This highlights the need for tailored clone detection and refactoring tools to maintain code consistency and software quality in quantum software.

\section{Future Plans}\label{sec:future-plans}
Building on our initial findings, we plan to extend this study in several directions to deepen our understanding of code clones in quantum software:

\begin{itemize}[leftmargin=1.3em]
\setlength{\itemsep}{1.0pt}
    \item \textit{Developing quantum-specific clone detection tools:} We will design tools for quantum software, focusing on more accurately detecting Type-2 and Type-3 clones. Current methods are limited in addressing the structural complexities of quantum programs, and improved detection algorithms can significantly enhance maintainability.

    \item \textit{Analyzing the impact of clones on scalability:} Future research will explore the impact of code clones on the performance and scalability of larger quantum programs, particularly in hybrid quantum-classical systems. Understanding how clones influence software efficiency will be critical for developing scalable quantum applications.

    \item \textit{Exploring refactoring strategies:} We aim to develop strategies for automated refactoring and best practices to mitigate the risks of code clones. This includes adapting existing techniques to the quantum domain, thus improving the robustness of quantum software.

    \item \textit{Towards a full-length study:} These efforts will form the basis of a comprehensive study, expanding on our preliminary results. The goal is to provide deeper insights into the role of code clones in quantum software development.
    %and to contribute to advancing quantum software engineering practices as the field moves towards commercial viability.
\end{itemize}

%\section*{Acknowledgement} % not allowed in review
%This work is supported in part by JSPS KAKENHI Grant No. JP23H03372 and No. JP24K14908.

\bibliographystyle{IEEEtran}
\bibliography{reference}
\end{document}

%% file: macros.tex
\usepackage{times}
\usepackage{graphicx}
\usepackage{epsf}
\usepackage{verbatim}
\usepackage{url}
\usepackage{color}
\usepackage{alltt}

\newcommand{\Comment}[1]{}

\newcommand{\SmallSpace}{\vspace*{-1.4ex}}